\documentclass[
twocolumn,
amsmath,
amssymb,
aps,
pra,
superscriptaddress]{revtex4-2}

\usepackage{xcolor}
\usepackage{graphicx}
\usepackage{dcolumn}
\usepackage{bm}
\usepackage{float}
\usepackage{upgreek}
\usepackage{hyperref}
\usepackage[capitalize]{cleveref}
\usepackage{physics}
\usepackage{siunitx}
\usepackage{dsfont}
\usepackage{makecell}
\usepackage{multirow}

\begin{document}

\title{Deterministic Generation of Linear Photonic Cluster States with Semiconductor Quantum Dots: A Detailed Comparison of Different Schemes}

\author{Nikolas Köcher}
\affiliation{Department of Physics and Center for Optoelectronics and Photonics Paderborn (CeOPP), Paderborn University, Warburger Straße 100, 33098 Paderborn, Germany}
\affiliation{Institute for Photonic Quantum Systems (PhoQS), Paderborn University, Warburger Straße 100, 33098 Paderborn, Germany}

\author{Stefan Schumacher}
\affiliation{Department of Physics and Center for Optoelectronics and Photonics Paderborn (CeOPP), Paderborn University, Warburger Straße 100, 33098 Paderborn, Germany}
\affiliation{Institute for Photonic Quantum Systems (PhoQS), Paderborn University, Warburger Straße 100, 33098 Paderborn, Germany}
\affiliation{Wyant College of Optical Sciences, University of Arizona, Tucson, AZ 85721, USA}

\date{\today}

\begin{abstract}
Photonic graph states are key resource states for measurement based quantum information processing. As semiconductor quantum dots are excellent deterministic photon emitters, several protocols using them for the generation of linear cluster states have been proposed, either based on constant precession of a hole or electron spin in a weak magnetic field, or based on optical spin control, in a stronger magnetic field. We theoretically compare four such schemes, using polarization or time-bin encoding, respectively, for a range of cavity environments and spin coherence times. In particular we study how different error mechanisms affect the different schemes, using a microscopic model of the spin control, the excitation and emission dynamics, and of the phonon bath. We find the spin-precession based schemes to scale well with strong cavity enhancement and to be naturally robust against phonon-induced decoherence, while the schemes using optical spin control can perform well for lower spin coherence times and are strongly dependent on the cooperativity of the cavity induced cycling transition. Our results provide a regime map for choosing between magnetic-field-driven and optically controlled protocols depending on spin coherence time, Purcell enhancement, and suppression of unwanted decay channels.

\end{abstract}

\maketitle

\section{Introduction}

\begin{figure}[b]
    \centering
    \includegraphics[width=1.0\columnwidth]{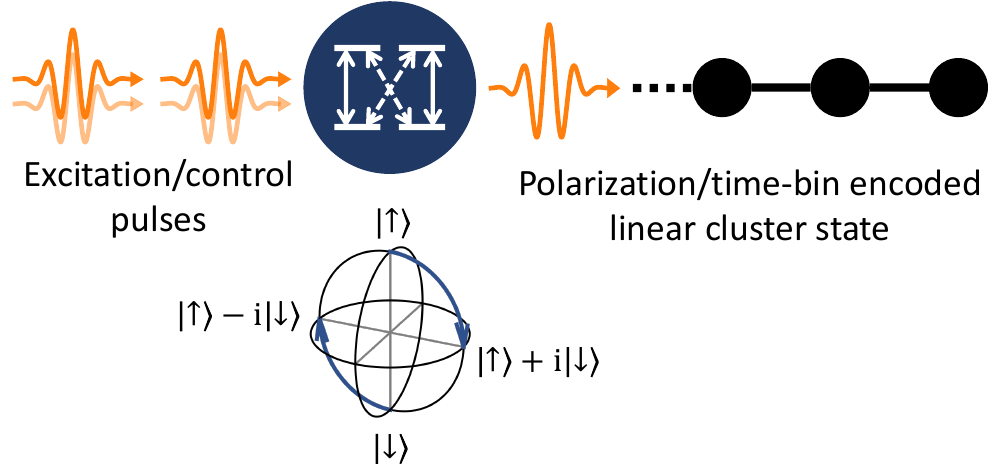}
    \caption{Schematic of the cluster state generation process. Periodic driving of the emitter, interspersed with qubit rotations, leads to the generation of an entangled string of either polarization or time-bin entangled photons.}
    \label{fig:temp0}
\end{figure}

Graph states are a type of multipartite entangled state with applications for quantum computation \cite{briegel2001persistent, raussendorf2001one, raussendorf2003measurement, raussendorf2007topological, briegel2009measurement, paesani2023high} and quantum communication \cite{zwerger2012measurement, azuma2015all}. Photonic graph states can be generated by combining single-photon sources with probabilistic entangling gates \cite{li2020multiphoton, istrati2020sequential, pont2024high, chen2024heralded} or by fusion of entangled photon pairs from spontaneous parametric down-conversion sources \cite{walther2005experimental, zhang2006experimental, lu2007experimental}. However, the probabilistic nature of these approaches limits the scalability for higher numbers of photonic qubits. Photonic graph states can instead be directly generated deterministically by using quantum emitters with a suitable level structure, as sketched in figure \ref{fig:temp0}. Alternatively, larger graph states can be built by fusing subgraphs that are generated deterministically, substantially reducing the number of required successful fusions \cite{manohar2026protocol, chan2026practical}.

A specific subset of graph states, so called linear cluster states, can be generated with a single quantum emitter. One of the most promising platforms for deterministic photon emitters are semiconductor quantum dots (QDs), due to their high brightness and indistinguishability of the photons \cite{michler2000quantum, wei2014deterministic, kuhlmann2015transform, ding2016demand, somaschi2016near, wang2016near, heindel2017bright, schweickert2018demand, huber2018strain, liu2018high, reindl2019highly, uppu2020scalable, tomm2021bright, thomas2021bright, jonas2022nonlinear, pelucchi2022potential, ding2025high, baltisberger2025indistinguishable, heinisch2026high}. The first scheme for the generation of linear photonic cluster states with quantum dots was proposed by Lindner and Rudolph (LR) and uses a spin in a singly charged quantum dot \cite{lindner2009proposal}. A weak constant in-plane magnetic field leads to precession of the spin and appropriately timed periodic excitation of the corresponding trions leads to the emission of a chain of photons entangled in their polarization degree of freedom. Linear cluster states of several photons have been generated with this approach, using either an electron \cite{coste2023high, huet2025deterministic} or a hole \cite{cogan2023deterministic, su2024continuous, laccotripes2025entangled} as the spin qubit. A similar scheme has also been demonstrated using a dark exciton \cite{schwartz2016deterministic}. Unequal g-factors of the ground state spin and the trion generally cause errors and lower the achievable fidelity, but can be partially mitigated by cavity enhancement of the trion decay rate. Alternatively, schemes based on optical spin control have been proposed and demonstrated \cite{lee2019quantum, tiurev2022high, appel2022entangling, meng2024deterministic}, which avoid these kinds of errors, but they require an artificial cavity-induced cycling transition and are only compatible with the emission of time-bin encoded cluster states \cite{vezvaee2022deterministic}.

In the present work, we compare the performance of these two approaches for a singly positively charged semiconductor quantum dot in an optical microcavity. We consider a total of four schemes: the originally proposed LR scheme, a version of the LR scheme modified to emit time-bin encoded cluster states and two schemes using optical spin control, either via a single ultrafast pulse or via Raman pulses.

The present work is structured as follows: we first introduce the generation schemes and our model of the quantum-dot-cavity system in detail. We then describe our approach for efficiently tracking the fidelity of the polarization and time-bin encoded cluster states using photonic correlation functions. Using this, we compare the different schemes in dependence of the hole spin coherence time and the cavity environment. We show that the optimal choice of scheme strongly depends on these parameters and that the spin-precession based schemes scale well with increasing coherence times, while the scheme based on ultrafast optical control performs well for low coherence times. We also show that the schemes using an artificially induced cycling transition strongly benefit from suppression of unwanted rivaling transitions. We additionally consider the influence of the phononic environment on the different state preparation schemes and show that the original LR scheme using polarization encoding is intrinsically immune to phonon-induced decoherence.

\section{Cluster State Generation Schemes and Model System}

\begin{figure*}[t]
    \centering
    \includegraphics[width=1.0\textwidth]{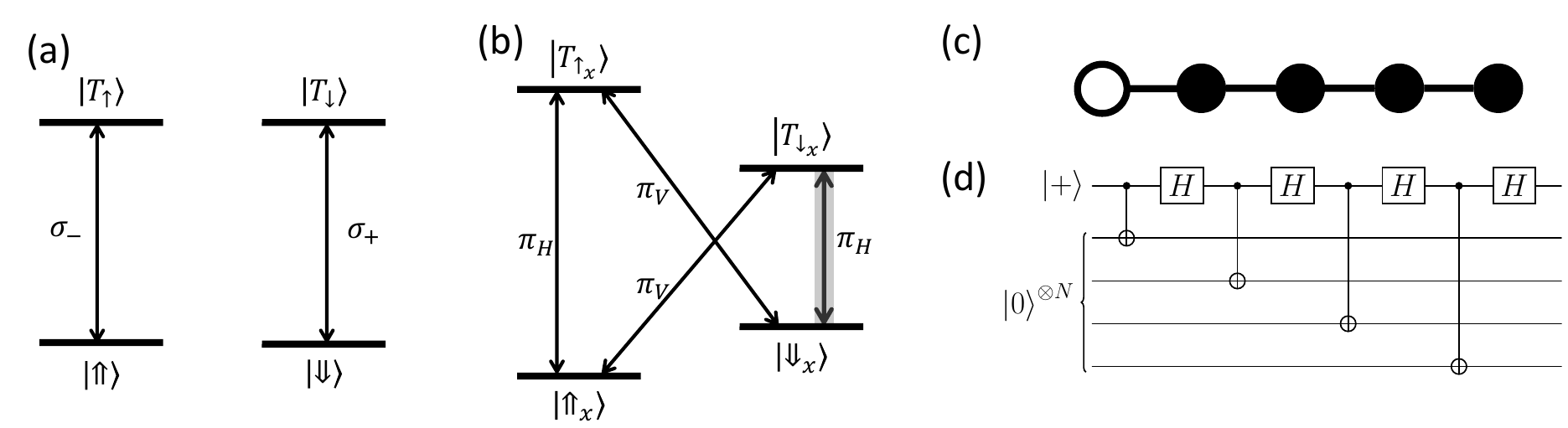}
    \caption{(a) Level scheme and selection rules for a singly positively charged QD without an external magnetic field. (b) Selection rules for the same QD, but with an external in-plane magnetic field $B_x$. The transition $\ket{\Downarrow_x}\leftrightarrow\ket{T_{\downarrow_x}}$ marked in light gray is selectively cavity enhanced. (c) Graphical depiction of a five qubit linear cluster state, obtained after four cycles of entangled photon generation. The first vertex represents the emitter, while the solid vertices each represent a photonic qubit, which may be polarization or time-bin encoded. A $Z$-measurement on the emitter removes its vertex from the graph and leaves a purely photonic cluster state. (d) Circuit diagram representation of the linear cluster state generation protocols, with the first qubit being the emitter and following ones being the photonic qubits. The photon emission step is equivalent to a CNOT gate on the emitter (control) and the newly created photon (target). The Hadamard gates $H$ may be replaced by $\pi/2$-rotations around an arbitrary axis in the xy-plane of the Bloch sphere.}
    \label{fig:temp1}
\end{figure*}

In this section we describe the cluster state generation schemes and our model of the system. The full Hamiltonian of the system and the equations of motion can be found in Appendix \ref{sec:hamiltonian}.

The LR scheme for cluster state generation makes use of the degenerate ground states and a pair of excited states of a singly charged QD, as depicted in figure \ref{fig:temp1} (a). We consider a positively charged QD, as holes generally have longer coherence times $T_2^*$ than electrons, but the model can easily be adapted to electrons. The ground states of our system are the two spin projections of a heavy hole along the optical axis $\ket{\Uparrow}$ and $\ket{\Downarrow}$, which we label the z-axis. These states optically couple to a pair of trion states ${\ket{T_\uparrow}=\ket{\Uparrow\Downarrow\uparrow}}$ and ${\ket{T_\downarrow}=\ket{\Uparrow\Downarrow\downarrow}}$, each consisting of two heavy holes and an electron, via circularly polarized light. Unless mentioned otherwise, the trions have a natural lifetime of ${\gamma_\mathrm{rad}^{-1}=\SI{1}{\nano\s}}$. A photonic microcavity resonant with the bare hole-trion transitions with dot-cavity coupling strength $g$ and cavity emission rate $\kappa$ is used to reduce the effective trion lifetime $T_1$. A weak magnetic field $B_x$ applied in Voigt configuration couples $\ket{\Uparrow}$ and $\ket{\Downarrow}$ and leads to precession of the hole spin on a nanosecond time scale with a period of ${T_B=2\pi\hbar/(\mu_B g_\mathrm{hole}B_x)}$. We assume Landé factors of $g_h=-0.2$ for the hole and $g_t=0.4$ for the trion, which are close to experimentally measured values e.g. for \mbox{InGaAs/GaAs} \cite{prechtel2015electrically} or \mbox{InAs/GaAs} \cite{cogan2022spin} QDs.

\begin{figure*}[t]
    \centering
    \begin{tabular}{|c||c|c|c|c|}
         \hline
         Scheme & B-pol (1) & B-time (2) & UF-time (3) & Raman-time (4) \\
         \hline
         \multirow{2}{*}{Qubit gates} & \multicolumn{2}{c|}{\makecell{\includegraphics[width=0.15\linewidth]{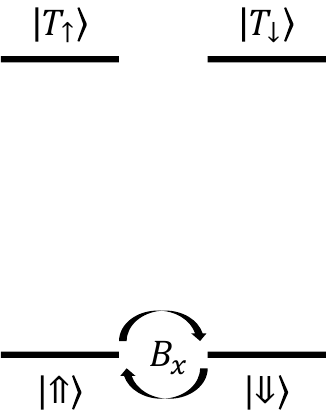}}} & \makecell{\includegraphics[width=0.15\linewidth]{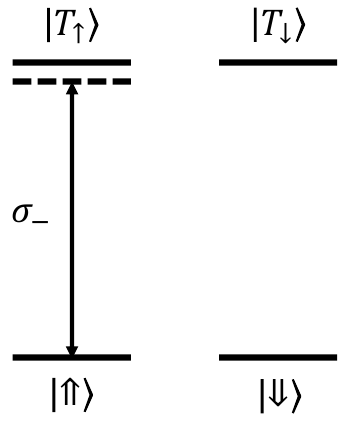}} & \makecell{\includegraphics[width=0.165\linewidth]{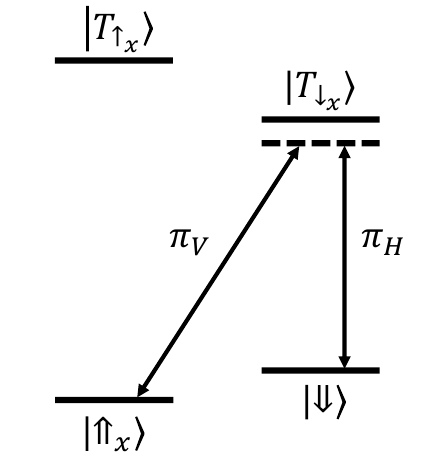}} \\
         \cline{2-5}
         & \multicolumn{2}{c|}{Weak magnetic field} & Fast pulse ($\sigma_-$) & Raman pulses ($\pi_H+\pi_V$) \\
         \hline
         \multirow{2}{*}{Photon emission} & \makecell{\includegraphics[width=0.15\linewidth]{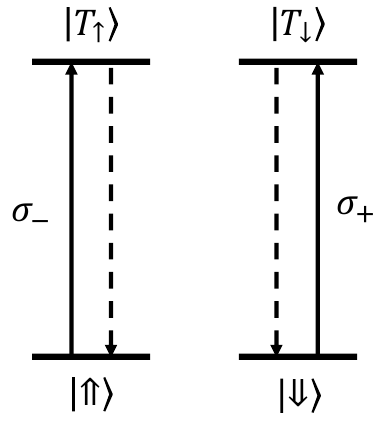}} & \makecell{\includegraphics[width=0.145\linewidth]{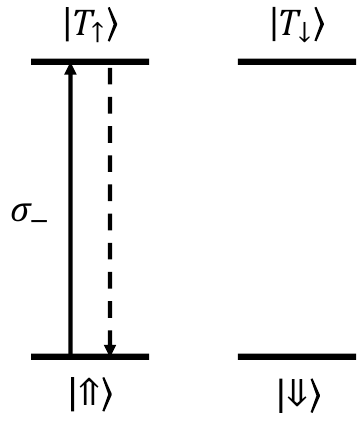}} & \multicolumn{2}{c|}{\makecell{\includegraphics[width=0.165\linewidth]{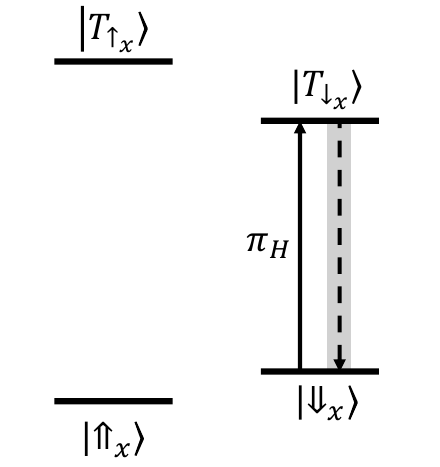}}} \\
         \cline{2-5}
         & Polarization encoded & \multicolumn{3}{c|}{Time-bin encoded} \\
         \hline
    \end{tabular}
    \caption{Overview of the four linear cluster state generation schemes we investigate: The original LR scheme using spin precession and polarization encoding (1), the LR scheme adapted for time-bin encoding (2) and two schemes using time-bin encoding and optical spin control via an ultrafast pulse (3) or via Raman pulses (4).}
    \label{fig:temp2}
\end{figure*}

This system enables the generation of linear cluster states with the following protocol: we initialize the QD in an equal superposition of the hole states $(\ket{\Uparrow}+\ket{\Downarrow})/\sqrt{2}$ and periodically excite both trions with linearly polarized pulses (${\pi_H \sim \sigma_+ + \mathrm{i}\sigma_-}$), which leads to the emission of photons, whose circular polarization is conditioned on the holes spin state. The timing of the pulses is chosen such that a quarter of a full spin precession period lies between the excitations, effectively realizing an $R_x(\pi/2)$-gate. Figure \ref{fig:temp2} shows a sketch of the spin control and photon emission steps. Note that while we start with an already initialized spin, the spin can be initialized by optical pumping or by adding one protocol cycle to effectively perform a $Z$-measurement of the hole spin at the start of the pulse sequence. Figure \ref{fig:temp1} (d) depicts this protocol as a quantum circuit for a 4 photon linear cluster state, where the uppermost qubit is the hole spin and the remaining qubits are the emitted photons. The trion excitation and photon emission step is equivalent to a controlled NOT gate between the hole spin (control) and the emitted photon (target). Figure \ref{fig:temp1} (c) shows a graphical representation of the resulting cluster state comprising the spin and 4 photons. A final $Z$-measurement of the spin can be used to decouple it from the rest of the state, resulting in a purely photonic cluster state. For simplicity we use resonant optical excitation of the trion states. In this context we note, that experimental implementations of this scheme use phonon-assisted excitation or excite higher-energy trions instead to spectrally separate the excitation pulses from the emitted photons, but this is still appropriately described by our model, as long as the excited trion lifetime is sufficiently short \cite{coste2023high, huet2025deterministic, cogan2023deterministic, su2024continuous, laccotripes2025entangled}.

An alternative to encoding the photonic qubits in the polarization degree of freedom of the emitted photons is time-bin encoding, which is more resilient to environmental influences when using fiber-based transmission \cite{wang2026time}. For time-bin encoding, the emission time of a photon in either an early time-bin $\ket{E}$ or a late time-bin $\ket{L}$ is used to encode the logical basis states of a qubit. The LR scheme can easily be modified to emit time-bin encoded cluster states by replacing the photon emission step with two excitations of the ${\ket{\Uparrow}\leftrightarrow\ket{T_\uparrow}}$ transition separated by an $R_x(\pi)$-gate. Figure \ref{fig:temp2} shows a sketch of modified excitation and photon emission step. The $R_x(\pi)$ gate is again implemented by waiting for half a spin precession period. This leads to the emission of a $\sigma_-$-photon, whose emission time is now conditioned on the initial state of the spin qubit and encodes a logical qubit of the cluster state.

A major source of infidelity for these cluster state generation schemes based on spin precession is the always-on magnetic field, leading to precession of the trion states. Unequal Landé factors of the hole and trion cause errors and generally reduce the achievable fidelity of the emitted cluster state. These errors can be reduced by reducing the ratio of trion lifetime to spin precession time via a cavity or by using weaker magnetic fields. The former approach requires high Purcell factor cavities, while the latter approach is limited by the finite coherence time of the hole spin.

An alternative approach for the generation of cluster states with the same QD-cavity system uses optical control of the hole spin, avoiding the errors associated with the always-on spin precession. The $\Lambda$-level structure required for this can be found when considering the eigenstates of our system including the Voigt configuration magnetic field. We refer to this basis as the x-basis and to the previously used basis as the z-basis. Figure \ref{fig:temp1} (b) shows the resulting level structure with the hole eigenstates
\begin{equation}
    \ket{\Uparrow_x/\Downarrow_x} = \frac{\ket{\Uparrow}\pm\ket{\Downarrow}}{\sqrt{2}}
\end{equation}
and trion eigenstates
\begin{equation}
    \ket{T_{\uparrow_x}/T_{\downarrow_x}} = \frac{\ket{T_\uparrow}\pm\ket{T_\downarrow}}{\sqrt{2}} .
\end{equation}
To allow for selective optical driving of the individual optical transitions, we use a strong magnetic field $B_x$ of several Tesla, lifting the degeneracy of the hole and of the trion states. We consider two optical driving schemes for spin control, both implementing z-rotations, which enable the generation of cluster states when combined with photon emission conditioned on the x-basis states. Figure \ref{fig:temp2} includes a sketch of these driving schemes. For the first scheme, we drive the ${\ket{\Uparrow} \leftrightarrow \ket{T_\uparrow}}$ transition with an ultrafast transitionless sech pulse with pulse area ${\alpha=2\pi}$, with detuning $\Delta$ and with a temporal pulse width of $\sigma$. For consistency we also use a sech shape for all other pulses. After the pulse has passed, the initial population is returned to the state $\ket{\Uparrow}$, but acquires a phase of ${\phi=\arctan{(\frac{\hbar}{\sigma\Delta})}}$, realizing an $R_z(\phi)$ gate on the hole spin qubit \cite{economou2006proposal}. To achieve high gate fidelities, the pulse is required to be fast compared to the spin precession time. For the second control scheme, Raman spin control, we drive both transitions from the ground states to the lower energy trion state $\ket{T_\downarrow}$ with Rabi frequencies $\Omega_H$ and $\Omega_V$ and with equal detuning $\Delta$. If the Rabi frequencies are sufficiently small compared to the detuning, this results in an effective coupling of ${\Omega_\mathrm{eff}=\Omega_H \Omega_V/(2\Delta)}$ between the ground states, implementing z-rotations $R_z(\phi)$ on the spin qubit \cite{chen2004theory, press2008complete}. The angle of rotation is determined by the total effective pulse area ${\int \Omega_\mathrm{eff}(t)dt}$. In this study, we always set ${\Omega_H=\Omega_V}$.

In the x-basis, the system lacks the intrinsic cycling transition(s) required for the generation of cluster states. Instead a nearly cycling transition has to be artificially induced via selective cavity-enhancement. In our case, we tune the cavities $\pi_H$-mode to be resonant with the ${\ket{\Downarrow_x} \leftrightarrow \ket{T_{\downarrow_x}}}$ transition. To limit unwanted enhancement of the ${\ket{\Uparrow_x} \leftrightarrow \ket{T_{\downarrow_x}}}$ transition, we detune the $\pi_V$-mode by $\Delta_\mathrm{cav}$, which can be achieved by for example using elliptical micropillar cavities, elliptical bullseye resonators \cite{wang2019towards, barbiero2024polarization} or photonic crystal waveguides \cite{tiurev2022high, appel2022entangling, meng2024deterministic}. The achieved enhancement is characterized by the ratio of the effective decay rate via the intended transition $\Gamma_H$ versus via the unwanted transition $\Gamma_V$, also called the cooperativity ${\mathcal{C}=\Gamma_H/\Gamma_V}$. The polarization selection rules of the system do not allow for a cycling transition with orthogonal polarization for $\ket{\Uparrow_x}$ to exist simultaneously, making it incompatible with polarization encoding. As we use $\ket{\Downarrow_x}$ as the ground state of the cycling transition, we now initialize the QD in an equal superposition of the x-basis hole eigenstates ${\ket{\Uparrow}=(\ket{\Uparrow_x}+\ket{\Downarrow_x})/\sqrt{2}}$, before performing the protocol cycles of alternating $R_z(\pi)$- and $R_z(\pi)$-rotations interspersed with the resonant excitation of ${\ket{\Downarrow_x} \leftrightarrow \ket{T_{\downarrow_x}}}$ and the corresponding photon emission.

Figure \ref{fig:temp2} shows an overview of the transition(s) used for photon emission and of the method of spin control for the four schemes we investigate in the following sections: the original LR scheme (B-pol), the LR scheme adapted to time-bin encoding (B-time) and two schemes using time-bin encoding and optical spin control, either via an ultrafast pulse (UF-time) or via Raman pulses (Raman-time).

\section{Photonic Correlations}\label{sec:correlationFunctions}

In this section, we detail our approach for efficiently assessing the fidelity of the emitted polarization and time-bin encoded cluster states. For this we make use of the stabilizer formalism, which enables us to characterize the cluster states using only three-photon correlations, as the numerical effort required for the evaluation of $N^\mathrm{th}$ order correlations scales exponentially with $N$. As a type of graph state, an $N$-qubit linear cluster state is defined as the simultaneous (+1)-eigenstate of its $N$ stabilizer generators
\begin{equation}
    \begin{split}
        S_1 &= X^{(1)} Z^{(2)}, \\
        S_i &= Z^{(i-1)}X^{(i)}Z^{(i+1)}, \ 2\leq i \leq N-1, \\
        S_N &= Z^{(N-1)}X^{(N)}.
    \end{split}
\end{equation}
Here $X^{(i)}$, $Y^{(i)}$ and $Z^{(i)}$ denote the Pauli matrices acting on the $i^\mathrm{th}$ emitted photon. We use the expectation values of the stabilizer generators as our primary figure of merit, where expectation values of ${\langle S_i \rangle=1}$ indicate an ideal cluster state.

We track the emitted photons via photonic correlation functions. The relevant expectation values can be expressed as
\begin{equation}
    \langle X^{(1)} Z^{(2)} \rangle = \frac{2 \, \mathrm{Re}\{ \bar{g}_{0001}^{(2)}-\bar{g}_{0111}^{(2)} \}}{\sum_{i,j \in \{0, 1\}} \bar{g}_{ijji}^{(2)}}
\end{equation}
and
\begin{equation}
    \begin{split}
        &\langle Z^{(1)} X^{(2)} Z^{(3)} \rangle \\ 
        = &\frac{ 2 \, \mathrm{Re}\{ \bar{g}_{000010}^{(3)}-\bar{g}_{001110}^{(3)}-\bar{g}_{100011}^{(3)}+\bar{g}_{101111}^{(3)} \} }{\sum_{i,j,k \in \{0, 1\}} \bar{g}_{ijkkji}^{(3)}}
    \end{split}
\end{equation}
with the time integrated second and third order correlation functions
\begin{equation}\label{eq:g2_bar}
    \bar{g}_{ijkl}^{(2)} = \int \int dt_1 dt_2 \langle a_i^\dagger(t_1) a_j^\dagger(t_2) a_k(t_2) a_l(t_1) \rangle
\end{equation}
and
\begin{equation}\label{eq:g3_bar}
    \begin{split}
        &\bar{g}_{ijklmn}^{(3)} \\
        = &\int \int \int dt_1 dt_2 dt_3 \langle a_i^\dagger(t_1) a_j^\dagger(t_2) a_k^\dagger(t_3) a_l(t_3) a_m(t_2) a_n(t_1) \rangle.
    \end{split}
\end{equation}
The expectation values $\langle Y^{(1)} Z^{(2)} \rangle$ and $\langle Z^{(1)} Y^{(2)} Z^{(3)} \rangle$ can be obtained by taking the imaginary instead of the real part of the right hand sides of equations (\ref{eq:g2_bar}) and (\ref{eq:g3_bar}). For polarization encoding, the corresponding annihilation and creation operators $a^{(\dagger)}_i$ are the operators for the circularly polarized cavity modes
\begin{equation}
    \begin{split}
        a_0^{(\dagger)}(t)&=a_{\sigma_+}^{(\dagger)}(t), \\
        a_1^{(\dagger)}(t)&=a_{\sigma_-}^{(\dagger)}(t)
    \end{split}
\end{equation}
and the integration bounds are chosen such that $t_i$ fully captures the emission of the $i^\mathrm{th}$ photon. Later three-photon expectation values $\langle Z^{(n-1)} X^{(n)} Z^{(n+1)} \rangle$ are obtained by shifting the integration bounds. For time-bin encoding, the creation and annihilation operators correspond to the early and late time-bin of the same polarization mode with
\begin{equation}
    \begin{split}
        a_0^{(\dagger)}(t)&=a_E^{(\dagger)}(t)=a^{(\dagger)}(t), \\
        a_1^{(\dagger)}(t)&=a_L^{(\dagger)}(t)=a^{(\dagger)}(t+T),
    \end{split}
\end{equation}
where $T$ is the separation between the start of the early and the start of the late time-bin and both bins have the same length \cite{bauch2024time, bracht2024theory}. The integration bounds are chosen such that $t_i$ fully captures the emission in the $i^\mathrm{th}$ early bin, so that $t_i+T$ captures the photon(s) in the late bin. In Appendix \ref{sec:correlationFunctionsExamples} we give a more detailed example for the third order correlation function $\bar{g}^{(3)}_{011100}$ for all four schemes.

We note that the schemes as described in the previous section may generate states that are not exactly equal to linear cluster states, but are equivalent up to local $R_z$ gates on the individual photons. Depending on the scheme, this can be corrected by adjusting the direction of the in-plane magnetic field or the phases of the driving laser pulses. As we are primarily interested in the entanglement properties of the emitted states, we do not explicitly carry out these corrections. Instead of directly using $\langle X^{(1)} Z^{(2)} \rangle$ as our figure of merit, we compute ${\langle X^{(1)} Z^{(2)} \rangle + \mathrm{i} \langle Y^{(1)} Z^{(2)} \rangle}$ and use its modulus. This is justified, as local $R_z$ gates can always be applied to the photons, changing the phase of ${\langle X^{(1)} Z^{(2)} \rangle + \mathrm{i} \langle Y^{(1)} Z^{(2)} \rangle}$ such that the expectation value is real and positive. The same argument holds for ${\langle Z^{(1)} X^{(2)} Z^{(3)} \rangle + \mathrm{i} \langle Z^{(1)} Y^{(2)} Z^{(3)} \rangle}$ and all following stabilizer generators. In the following, we simply refer to the modulus of the complex valued expressions as $\langle X^{(1)} Z^{(2)} \rangle$ and $\langle Z^{(1)} X^{(2)} Z^{(3)} \rangle$.

To estimate the length of the cluster states that can be generated, we use the following inequality based on a witness detecting genuine $N$-partite entanglement in graph states \cite{toth2005entanglement}. For a linear cluster state, genuine $N$-partite entanglement can be detected for all $N$ that satisfy 
\begin{equation}\label{eq:entanglementLength}
    N < \frac{\langle X^{(1)} Z^{(2)} \rangle}{1-\langle Z^{(1)} X^{(2)} Z^{(3)} \rangle} + 1 .
\end{equation}
We refer to the right hand side of the inequality as the entanglement length $L$. Note that due to being based on an entanglement witness, there may be cases, where genuine $N$-partite entanglement of more than $L$ qubits exists, that our witness does not detect. A derivation of the inequality can be found in Appendix \ref{sec:lowerBound}.

We also note, that there exist similar lower bounds e.g. on the state fidelity and the localizable entanglement, which similarly make use of the expectation values of the stabilizer generators \cite{nutz2017proposal, cramer2010efficient}.

\section{Results}

In the first part of this section we compare the cluster state generation schemes in dependence of the cavity environment and the hole spin coherence time $T_2^*$. In the second part, we include coupling of the electronic states to phonons and investigate how this affects the different schemes.

\subsection{Comparing the Schemes}

Our procedure for fairly comparing the schemes is made up of two steps. For each scheme and combination of system parameters, we first find the length of the excitation and rotation pulses, the strength of the external magnetic field and the length of the time-bins that optimize $\langle X^{(1)} Z^{(2)} \rangle$. Table \ref{tab:parameters} shows which parameters are optimized and which are fixed for each scheme. We then calculate the computationally more expensive expectation value $\langle Z^{(1)} X^{(2)} Z^{(3)} \rangle$ based on the optimized parameters, which we use as the primary figure of merit. This is justified as all three-photon stabilizer generators $\langle Z^{(1)} X^{(2)} Z^{(3)} \rangle$ have the same expectation value. For some results, we additionally show the entanglement length introduced in the previous section to highlight the length of the generated cluster states.

In the following, we vary the hole spin coherence time $T_2^*$, the QD-cavity coupling strength $g$, the cavity emission rate $\kappa$ and the emission rate into non-cavity modes $\gamma_\mathrm{rad}$. The former depends on the type of spin used and the nuclear spin environment. The latter three can be tuned by engineering the cavity environment of the QD. 

\begin{table}[t]
    \centering
    \begin{tabular}{|c|c|c|c|}
         \hline
         Scheme & Parameter & Symbol & Value \\
         \hline %\hline
         \multirow{2}{*}{\shortstack{B-pol \\ (1)}} & Magnetic field & $B_1$ & opt. \\
         %\cline{2-4}
          & Ex. pulse length & $\sigma_{\mathrm{ex},1}$ & \SI{0.5}{\pico\s} \\
         \hline %\hline
         \multirow{2}{*}{\shortstack{B-time \\ (2)}} & Magnetic field & $B_2$ & opt. \\
         %\cline{2-4}
          & Ex. pulse length & $\sigma_{\mathrm{ex},2}$ & \SI{0.5}{\pico\s} \\
         \hline %\hline
         \multirow{5}{*}{\shortstack{UF-time \\ (3)}} & Magnetic field & $B_3$ & opt, $\leq\SI{9}{\tesla}$ \\
         %\cline{2-4}
          & Cav. V mode det. & $\Delta_{\mathrm{cav}, 3}$ & \SI{-1}{\meV} \\
         %\cline{2-4}
         & Ex. pulse length & $\sigma_{\mathrm{gate}, 3}$ & \SI{0.5}{\pico\s} \\
         %\cline{2-4}
          & Rot. pulse length & $\sigma_{\mathrm{ex}, 3}$ & opt. \\
         %\cline{2-4}
          & Bin length & $T_3$ & opt. \\
         \hline %\hline
         \multirow{6}{*}{\shortstack{Raman-time \\ (4)}} & Magnetic field & $B_4$ & \SI{9}{\tesla} \\
         %\cline{2-4}
          & Cav. V mode det. & $\Delta_{\mathrm{cav}, 4}$ & \SI{-1}{\meV} \\
         %\cline{2-4}
         & Ex. pulse length & $\sigma_{\mathrm{ex}, 4}$ & opt. \\
         %\cline{2-4}
          & Rot. pulse length & $\sigma_{\mathrm{gate}, 4}$ & opt. \\
         %\cline{2-4}
          & Rot. pulse det. & $\Delta_{\mathrm{gate}, 4}$ & \SI{-2}{\meV} \\
         %\cline{2-4}
          & Bin length & $T_4$ & $10(g^{-1}+\sigma_\mathrm{gate}^{-1})$ \\
         \hline
    \end{tabular}
    \caption{Tunable parameters for each of the generation schemes. The parameters marked with opt. are optimized for each set of system parameters ($T_2^*$, $\gamma_\mathrm{rad}$, $g$, $\kappa$) to allow for a fair comparison between the schemes.}
    \label{tab:parameters}
\end{table}

\begin{figure}[t]
    \centering
    \includegraphics[width=1.0\linewidth]{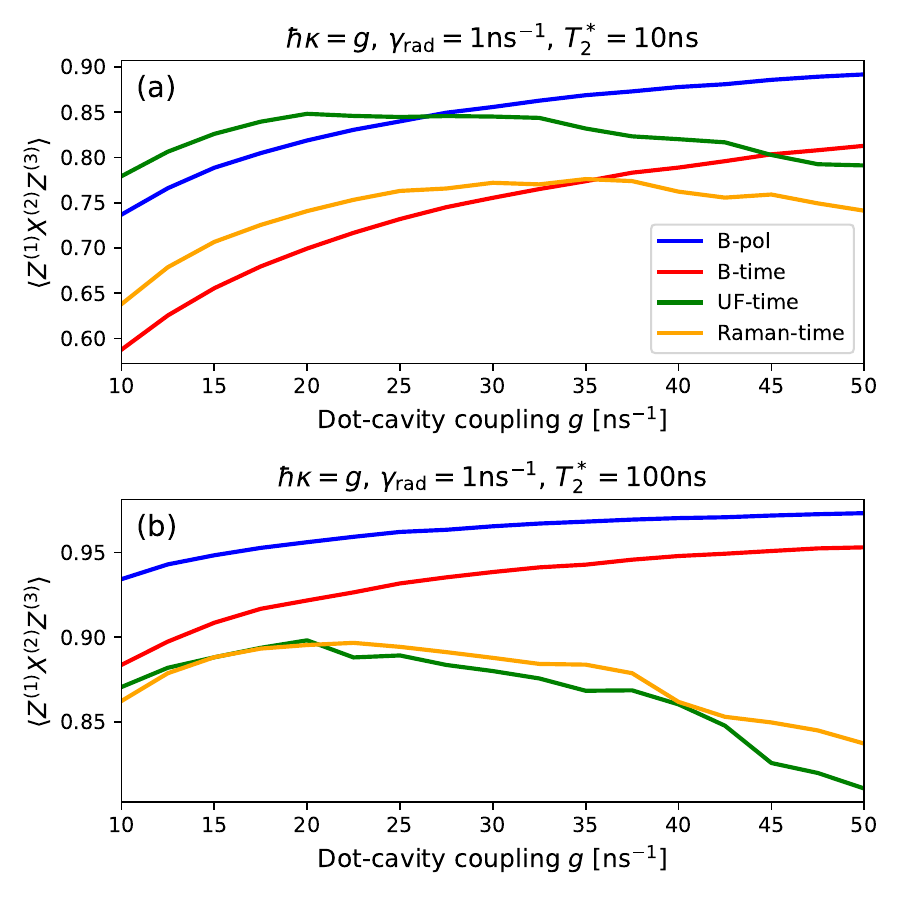}
    \caption{Comparison of $\langle Z^{(1)} X^{(2)} Z^{(3)} \rangle$ for the cluster state generation schemes in dependence of the dot-cavity coupling $g$ for different hole spin coherence times $T_2^*=\SI{10}{\nano\s}$ (a) and $T_2^*=\SI{100}{\nano\s}$ (b).}
    \label{fig:fig1}
\end{figure}

We begin our investigations by considering the influence of the QD-cavity coupling strength $g$ on the four schemes. Figure \ref{fig:fig1} (a) shows the expectation value of the first three-photon stabilizer generator in dependence of $g$ for a hole spin coherence time of ${T_2^*=\SI{10}{\nano\s}}$. We mainly use $g$ to tune the enhancement of the trion lifetime $T_1$. For the case of the spin-precession based schemes, the range of $g$ shown here corresponds to a reduction of the trion lifetime by a factor of approximately 5-25. The enhancement is lower for the schemes based on optical spin control due to the lower dipole moment of the x-basis transitions compared to the z-basis ones. The spin-precession based schemes strictly benefit from stronger QD-cavity coupling, as their fidelity is mainly determined by the ratios of trion lifetime, spin precession time and coherence time. High fidelities require
\begin{equation}\label{eq:optimalRatiosLR}
T_1 \ll T_B \ll T_2^*,
\end{equation}
as shorter trion lifetimes allow for faster gate times and thus less decoherence. Comparing the generation of polarization and time-bin encoded cluster states shows that here the former encoding method is clearly superior, as the latter requires twice the number of excitation pulses and three times as long to generate a single photonic qubit, assuming the same bin length. The schemes based on optical spin control initially also benefit from stronger QD-cavity coupling, because of the enhanced cooperativity of the cycling transition, but too strong coupling leads to a decrease in the cluster state fidelity. This can mostly be attributed to reexcitation caused by the finite length of the excitation pulses, which need to be sufficiently long to allow for the spectrally selective excitation of the ${\ket{\Downarrow_x}\rightarrow\ket{T_{\downarrow_x}}}$ transition. Reexcitation in the presence of strong cavity enhancement may be avoided  by using advanced excitation methods, such as the red-detuned swing-up scheme, but that lies outside of the scope of this work \cite{bracht2021swing, heinisch2024swing}. In principle, the increased width of the cavity resonance could also lead to unintended enhancement of rivaling hole-trion transitions, but because of the large magnetic field-induced splitting of the electronic states and the splitting of the cavity modes, this does not contribute significantly in our case. Comparing spin control via a single ultrafast pulse and Raman pulses favors the former due to the shorter gate time, which reduces the influence of decoherence. Finally, a comparison of the different types of schemes shows, that the scheme using ultrafast pulses for spin control performs best for weaker QD-cavity coupling, but is overtaken by the original LR scheme with polarization encoding for ${g\gtrsim\SI{20}{\per\nano\s}}$.

This crossing point between the optimal schemes generally depends on the other system parameters, like the hole spin coherence time. Figure \ref{fig:fig1} (b) again shows $\langle Z^{(1)} X^{(2)} Z^{(3)} \rangle$ in dependence of $g$, but with the coherence time extended to ${T_2^*=\SI{100}{\nano\s}}$. The qualitative behavior of the individual schemes remains similar. While the fidelity generally increases for all schemes, the spin-precession based schemes benefit much more strongly from the increased coherence time due to the intrinsic link between the bin length and errors caused by trion precession, as described in equation \ref{eq:optimalRatiosLR}. For ${T_2^*=\SI{100}{\nano\s}}$ this makes the original LR scheme optimal for the whole range of $g$ depicted here. As opposed to the case with a shorter coherence time, we also find both schemes using optical spin control to reach similar fidelities, as the length of the Raman pulses becomes less detrimental for longer coherence times.

\begin{figure}[t]
    \centering
    \includegraphics[width=1.0\linewidth]{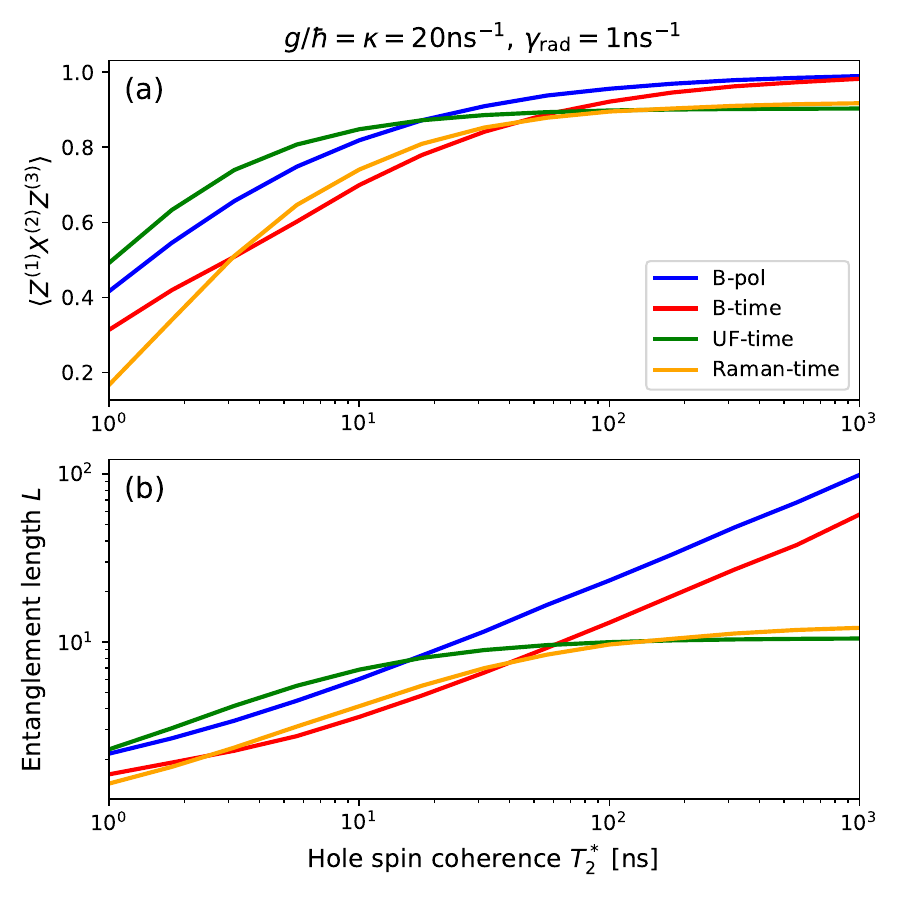}
    \caption{Comparison of $\langle Z^{(1)} X^{(2)} Z^{(3)} \rangle$ (a) and of the entanglement length $L$ (b) for the different schemes in dependence of the hole spin coherence time $T_2^*$.}
    \label{fig:fig1_}
\end{figure}

We further investigate the dependence on the hole spin coherence time. Figure \ref{fig:fig1_} (a) shows $\langle Z^{(1)} X^{(2)} Z^{(3)} \rangle$ for $g/\hbar=\SI{20}{\per\nano\s}$ in dependence of coherence times from \SI{1}{\nano\s} to \SI{1}{\micro\s}. As some schemes reach expectation values close to $1$, we additionally depict the entanglement length $L$ for the same set of system parameters in figure \ref{fig:fig1_} (b). Clearly, all four schemes benefit from increased coherence times, but they scale differently. For lower coherence times, the scheme using ultrafast pulses for spin control performs best due to the short gate times, despite using time-bin encoding, which requires two excitation and emission steps per photonic qubit. With longer coherence times, both schemes using optical spin control eventually reach plateaus significantly below $\langle Z^{(1)} X^{(2)} Z^{(3)} \rangle=1$, where they are primarily limited by the finite cooperativity of the cycling transition. For long coherence times, we again see Raman spin control outperform ultrafast pulses. The gate fidelity of the latter scheme could in principle be slightly improved by using a weaker magnetic field or faster pulses, but the former approach clashes with the requirements for spectrally selective trion excitation and we limit ourselves to pulses with $\sigma\ge\SI{0.5}{\pico\s}$. The main limitation remains the finite cooperativity. The spin-precession based schemes do not suffer from this drawback. While the condition $T_1 \ll T_B \ll T_2^*$ strongly limits the cluster state fidelity for short coherence times, long coherence times allow for long time-bins and very high fidelities. The polarization and time-bin versions of the LR scheme surpass the scheme using ultrafast pulses for $T_2^* \approx \SI{20}{\nano\s}$ and $T_2^* \approx \SI{50}{\nano\s}$ respectively. In our model, in the limit of vanishing decoherence the only factor limiting the cluster state fidelity is the finite duration of the excitation pulses. Hence the fidelity and the entanglement length scale well with increasing coherence times, even on the order of \SI{1}{\micro\s}. We note that especially for the spin-precession based schemes, this means sacrificing a higher repetition rate for a higher state fidelity.

\begin{figure}[t]
    \centering
    \includegraphics[width=1.0\linewidth]{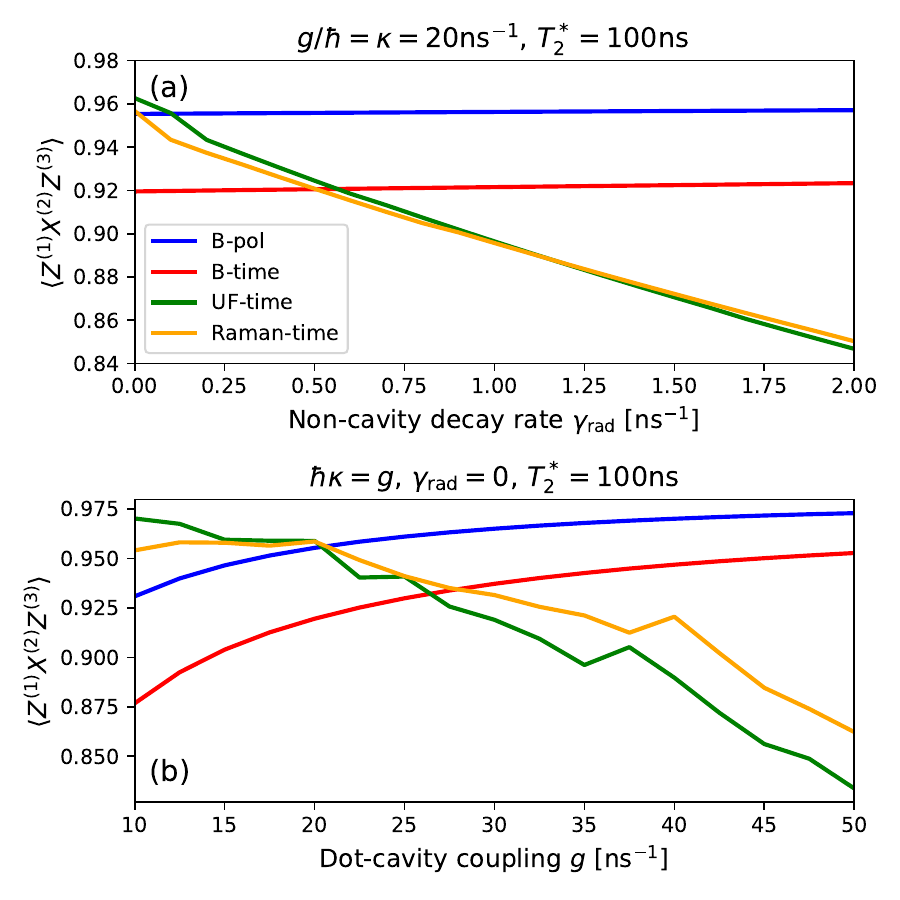}
    \caption{Influence of suppressing unwanted transitions. \mbox{(a) Comparison} of $\langle Z^{(1)} X^{(2)} Z^{(3)} \rangle$ for the different schemes in dependence of the non-cavity emission rate $\gamma_\mathrm{rad}$. $\gamma_\mathrm{rad}=0$ corresponds to perfect suppression, while $\gamma_\mathrm{rad}=\SI{1}{\per\nano\s}$ corresponds to the previously shown results. (b) The expectation value $\langle Z^{(1)} X^{(2)} Z^{(3)} \rangle$ in dependence of the dot-cavity coupling $g$ for perfect suppression.}
    \label{fig:fig2}
\end{figure}

An important approach for improving the cluster state generation schemes based on optical spin control is improving the cyclicity of the cavity induced cycling transition. However, the straightforward path of enhancing the effective trion decay rate $\Gamma_H$ leads to high reexcitation probabilities. Alternatively, a high cooperativity can also be achieved by suppression of the unwanted decay channel $\Gamma_V$, for example via a photonic crystal waveguide \cite{tiurev2022high, appel2022entangling, meng2024deterministic}. We model this by adjusting the (non-cavity) radiative decay rate $\gamma_\mathrm{rad}$. Figure \ref{fig:fig2} (a) shows $\langle Z^{(1)} X^{(2)} Z^{(3)} \rangle$ for the four schemes in dependence of $\gamma_\mathrm{rad}$ for $g/\hbar=\SI{20}{\per\nano\s}$ and $T_2^* = \SI{100}{\nano\s}$. The precession-based schemes are nearly unaffected by this, although they see a slight increase in the cluster state fidelity with increasing $\gamma_\mathrm{rad}$, due to the decrease in trion lifetime. The schemes using optical spin control are much more strongly affected. With decreasing $\gamma_\mathrm{rad}$, the improved cooperativity leads to significantly higher cluster state fidelities. Interestingly, the scheme using ultrafast pulses scales slightly better with decreasing $\gamma_\mathrm{rad}$ than the one using Raman spin control. We attribute this to the fact that the former scheme implements the $R_z(\pi)$ gates with resonant $2\pi$-pulses, which result in a temporary trion occupation, making the scheme more susceptible to trion decay.

We now investigate how the cluster state generation schemes compare when considering the improved cooperativity. Figure \ref{fig:fig2} (b) shows $\langle Z^{(1)} X^{(2)} Z^{(3)} \rangle$ in dependence of $g$ for ${T_2^*=\SI{100}{\nano\s}}$ for the ideal case where ${\gamma_\mathrm{rad}=0}$. Compared to figure \ref{fig:fig1} (b), the spin-precession based schemes are nearly unaffected by this. The schemes using the cavity-induced cycling transition, however, see a significant improvement in the cluster state fidelity. This effect is strongest for low $g$, where the cluster state fidelity was previously primarily limited by the lower cooperativity. This, combined with the high reexcitation probabilities for larger $g$ shifts the optimal QD-cavity coupling strength to ${g\approx\SI{10}{\per\nano\s}}$. Comparing the different kinds of schemes shows that schemes using optical control are clearly superior for weaker QD-cavity coupling and are only matched by the LR scheme for strong cavity enhancement of ${g/\hbar\approx\SI{50}{\per\nano\s}}$.

\begin{figure}[t]
    \centering
    \includegraphics[width=1.0\linewidth]{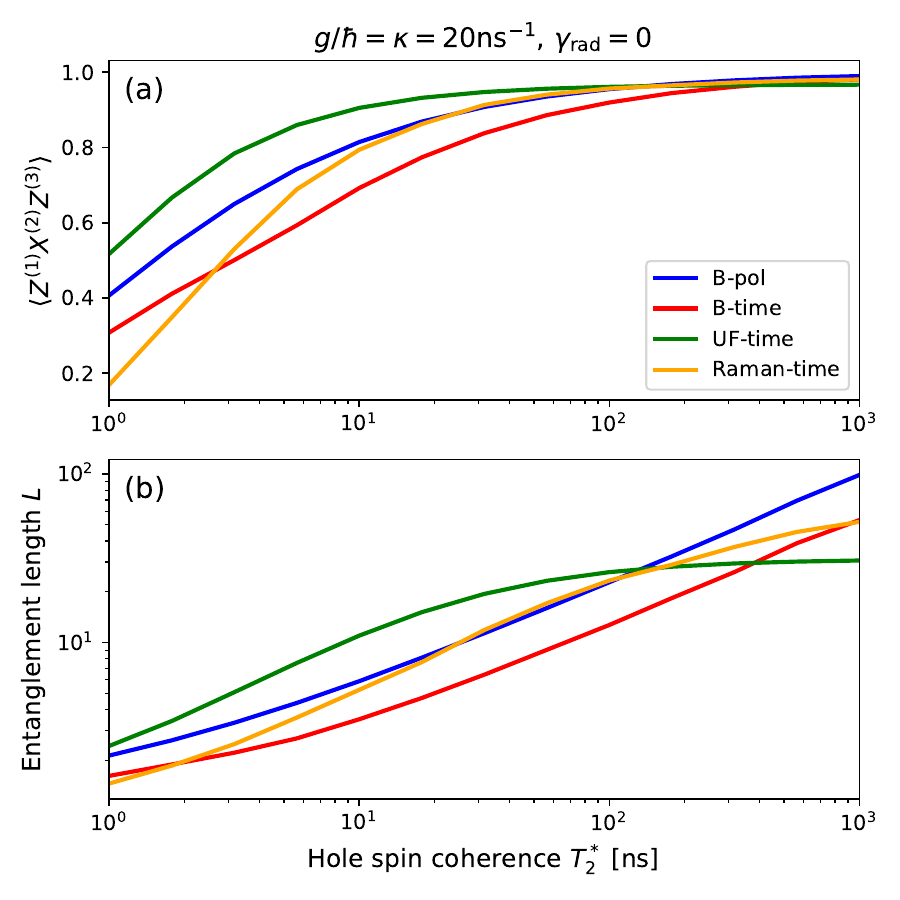}
    \caption{Comparison of $\langle Z^{(1)} X^{(2)} Z^{(3)} \rangle$ (a) and of the entanglement length $L$ (b) for the different schemes in dependence of the hole spin coherence time $T_2^*$ for the case of perfect suppression of unwanted transitions $\gamma_\mathrm{rad}=0$.}
    \label{fig:fig2_}
\end{figure}

The increased cooperativity significantly affects how the schemes scale with the hole spin coherence time. Figure \ref{fig:fig2_} (a) and (b) show $\langle Z^{(1)} X^{(2)} Z^{(3)} \rangle$ and the entanglement length $L$ in dependence of $T_2^*$ for $g/\hbar=\SI{20}{\per\nano\s}$ and $\gamma_\mathrm{rad}=0$. Compared to figure \ref{fig:fig1_} the spin-precession based schemes are again nearly unaffected by the change in $\gamma_\mathrm{rad}$, while the schemes based on optical spin control benefit from the enhanced cooperativity. The scheme using ultrafast pulses reaches an entanglement length of $L=30$ at $T_2^*=\SI{100}{\nano\s}$, but does not significantly improve for longer coherence times, as it is primarily limited by the gate fidelity, which is nearly independent of the spin coherence time. The scheme using Raman spin control on the other hand strongly benefits from the increased coherence time, even on the order of $T_2^*\approx\SI{1}{\micro\s}$, as it allows for longer Raman pulses and thus higher gate fidelities. Overall, the LR scheme is still optimal for longer coherence times $T_2^*\gtrsim\SI{100}{\nano\s}$, but suppressing the non-cavity emission allows for the generation of higher fidelity cluster states for lower coherence times.

\begin{figure}[t]
    \centering
    \includegraphics[width=1.0\linewidth]{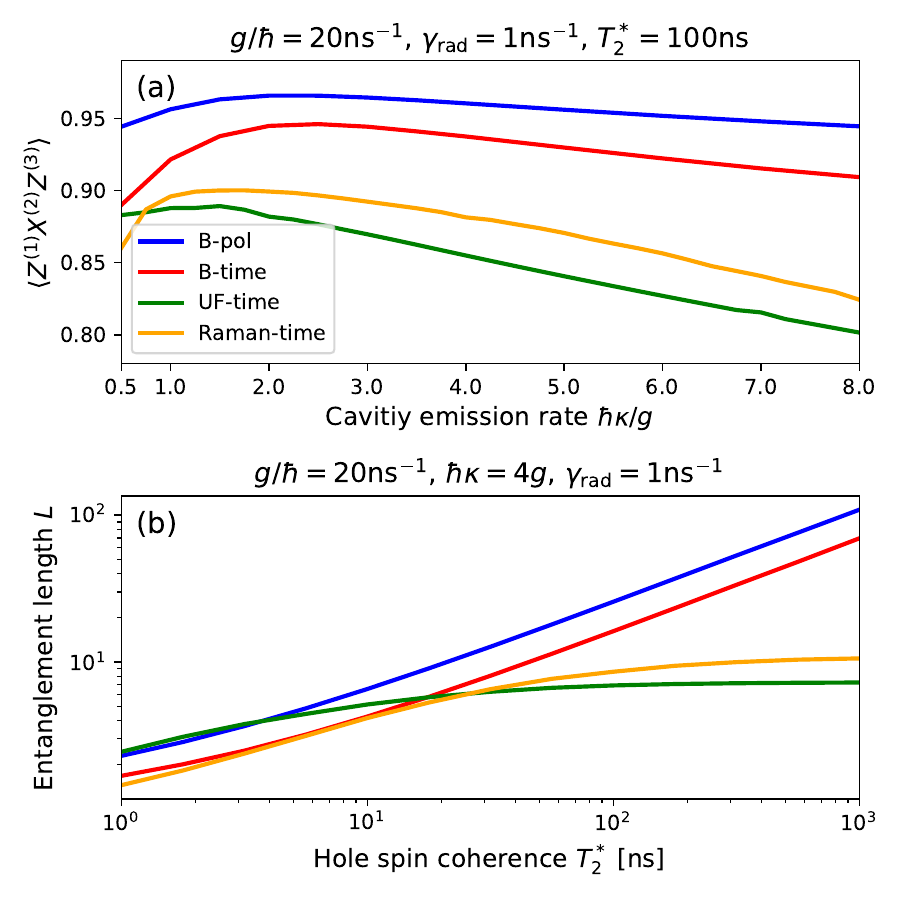}
    \caption{Comparison of $\langle Z^{(1)} X^{(2)} Z^{(3)} \rangle$ for the cluster state generation schemes in dependence of the cavity emission rate $\kappa$ for $g/\hbar=\SI{20}{\per\nano\s}$ (a) and in dependence of the dot cavity coupling $g$ for $\hbar\kappa=4g$ (b).}
    \label{fig:fig3}
\end{figure}

For the previous results we assumed a cavity emission rate of $\hbar\kappa=g$ and used the QD-cavity coupling strength primarily to tune the effective lifetime of the trions. We now investigate, whether the cavity emission rate influences the different generation schemes beyond the influence on the trion lifetime. Figure \ref{fig:fig3} (a) shows $\langle Z^{(1)} X^{(2)} Z^{(3)} \rangle$ in dependence of the cavity emission rate $\kappa$ for $g/\hbar=\SI{20}{\per\nano\s}$, $\gamma_\mathrm{rad}=\SI{1}{\per\nano\s}$ and $T_2^*=\SI{100}{\nano\s}$. The spin-precession based schemes reach a maximum fidelity for $\hbar\kappa\approx 2g$, which corresponds to the minimum of the trion lifetime. The same holds true for the schemes using the cavity-induced cycling transition, but as the x-basis transitions only see an effective QD-cavity coupling of $\tilde{g}=g/\sqrt{2}$, their maximum fidelity and minimum trion lifetime lie at $\hbar\kappa\approx2\tilde{g}=\sqrt{2}g$. We also find, that the scheme using Raman pulses falls off more quickly for $\hbar\kappa<g/\sqrt{2}$, while the scheme using ultrafast pulses falls off more quickly for $\hbar\kappa>g/\sqrt{2}$. As a further example, figure \ref{fig:fig3} (b) shows the entanglement length in dependence of $T_2^*$ for $g/\hbar=\SI{20}{\per\nano\s}$, $\hbar\kappa=4g$ and $\gamma_\mathrm{rad}=\SI{1}{\per\nano\s}$. A comparison with figure \ref{fig:fig1_} (b) shows, that the schemes using optical spin control are more strongly affected by the increase in $\kappa$, as $\hbar\kappa=4g$ is a stronger deviation from the ideal ratio. Nonetheless, the scaling with $T_2^*$ remains qualitatively similar for each scheme respectively and the change in fidelity is largely explained by the change in effective trion lifetime.

We note, that while our investigations focus on a hole spin, our predictions should also apply to electron spins in singly negatively charged QDs, which provide the same kind of level scheme as positively charged QDs. In the scope of our model only the g-factors and spin coherence times would need to be adjusted. To a lesser extend our model also applies to the case, where dark excitons are used as the entangling spin, but the natural fine structure splitting is not strong enough to allow for the implementation of optical spin control and a cavity induced cycling transition \cite{schwartz2016deterministic}.

We also note, that while we generally compare the different schemes for the same hole spin coherence time, the actual behavior of the coherence time may be more complex. The decoherence caused by e.g. the hyperfine interaction and charge noise behaves differently for electrons and holes \cite{cogan2022spin}, generally depends on the external magnetic field \cite{prechtel2016decoupling, huthmacher2018coherence}, and can be extended via spin cooling schemes \cite{meng2024deterministic, hogg2025fast, nguyen2023enhanced}. A rising coherence time for stronger magnetic fields might for example increase the fidelity of the schemes using optical spin control compared to the spin-precession based ones. The time-bin schemes might additionally benefit from a longer effective coherence time because of the spin echo effect caused by the $R_x(\pi)$ spin flips between the early and late time-bins \cite{lee2019quantum, tiurev2022high}.

\subsection{Influence of Phonons}\label{sec:phononInfluence}

In this section, we investigate the influence of phonons on the state preparation schemes. We include longitudinal acoustic phonons and deformation potential coupling, which is the dominant interaction mechanism at low temperatures \cite{quilter2015phonon, reiter2019distinctive}. As perturbative methods such as weak-coupling master equations, the polaron master equation or variational master equations fail at accurately describing strong coupling to phonons and strong driving simultaneously \cite{mccutcheon2010quantum, gustin2018pulsed,hanschkePRL2025}, we make use of a process tensor matrix product operator method \cite{cygorek2022simulation}. In Appendix \ref{sec:hamiltonian_phonons} we give details on the method and the Hamiltonian describing the phonons and their coupling to the electronic states. The interaction with the phonons is characterized by their spectral density
\begin{equation}
    J(\omega)=\alpha_p\omega^3\exp(-\frac{\omega^2}{2\omega_b^2}),
\end{equation}
where $\alpha_p$ is the phonon coupling strength and ${\hbar\omega_b=\SI{1}{\meV}}$ is the phonon cutoff frequency, which are based on realistic parameters for \mbox{InGaAs/GaAs} QDs \cite{quilter2015phonon}. We set the temperature to $T=\SI{4}{\kelvin}$.

\begin{figure}[t]
    \centering
    \includegraphics[width=1.0\linewidth]{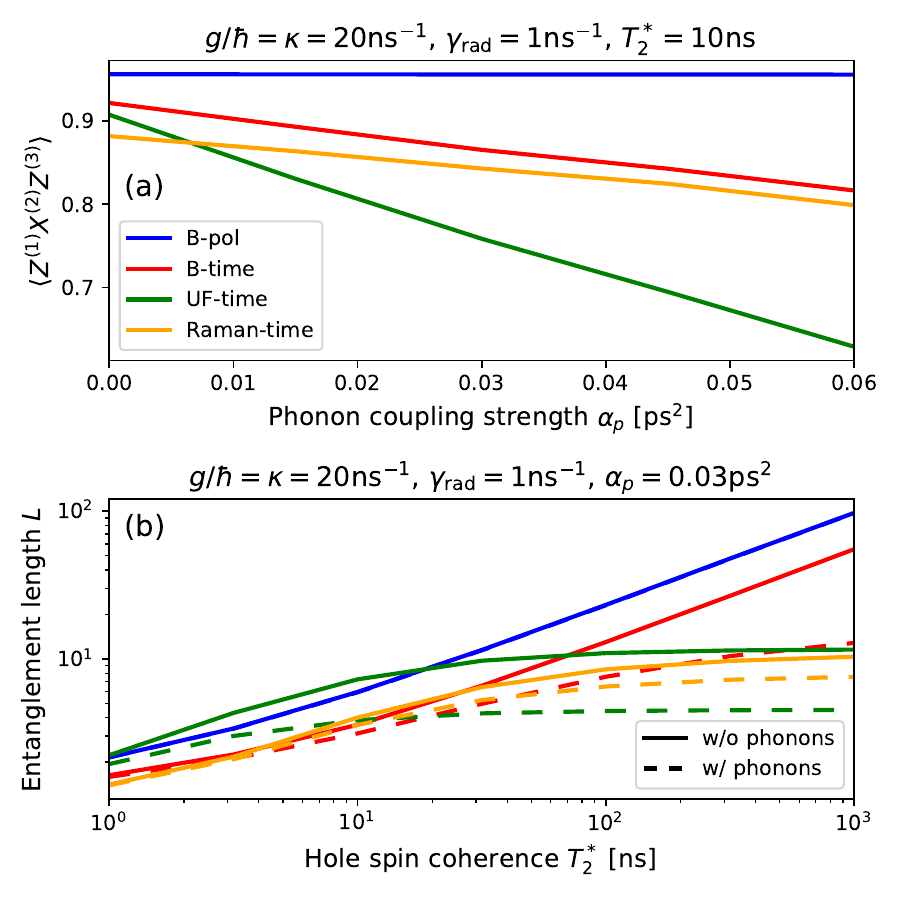}
    \caption{Influence of phonons on the cluster state generation. We include LA phonons and show $\langle Z^{(1)} X^{(2)} Z^{(3)} \rangle$ in dependence of the phonon coupling strength $\alpha_p$ (a). We also show $\langle Z^{(1)} X^{(2)} Z^{(3)} \rangle$ in dependence of the hole spin coherence time $T_2^*$ for the different schemes and compare the results obtained without and with the inclusion of phonons (${\alpha_p=\SI{0.03}{\pico\s\squared}}$) (b). Here, both lines for the original LR scheme are visually indistinguishable, as the schemes fidelity is nearly unaffected by coupling to phonons.}
    \label{fig:fig4}
\end{figure}

Figure \ref{fig:fig4} (a) shows $\langle Z^{(1)} X^{(2)} Z^{(3)} \rangle$ in dependence of the phonon coupling strength for the different schemes, where we use the optimized parameter sets from \mbox{figure \ref{fig:fig1} (a)} for $g/\hbar=\SI{20}{\per\nano\s}$. We find, that for all three time-bin-based schemes, phonon-induced decoherence results in lower cluster state fidelities for stronger phonon coupling strengths. We mostly avoid phonon-assisted excitation by only driving the system at or below the lowest energy transition of each respective polarization. We similarly avoid phonon-assisted cavity feeding by not populating the higher energy trion state. The exception to this are the $R_z(\pi)$ gates implemented with ultrafast pulses, which use a circularly polarized pulse resonant with the bare hole-trion transitions. This causes phonon-assisted excitation of the lower-energy trion and the increased sensitivity of this scheme to increasing phonon coupling. Interestingly, the original LR-scheme, which uses polarization encoding, is nearly unaffected be the coupling to phonons. This resistance against phonon-induced decoherence is caused by the symmetric excitation of both trions, which couple equally strongly to phonons. We further elaborate on this in Appendix \ref{sec:phononProtectionLR}. While the fidelity is unaffected by phonons for this scheme, all schemes experience a small decrease in brightness because of the renormalization of the Rabi frequency, but this can be mitigated by adjusting the pulse amplitudes.

In figure \ref{fig:fig4} (b) we show the entanglement length in dependence of the coherence time without and with (${\alpha_p=\SI{0.03}{\ps\squared}}$) coupling to phonons for ${g/\hbar=\kappa=\SI{20}{\per\nano\s}}$ and ${\gamma_\mathrm{rad}=\SI{1}{\nano\s}}$. For the original LR scheme, both lines nearly coincide, as the scheme is insensitive to coupling to phonons, while the scheme based on ultrafast control pulses is again affected most strongly. For the present parameter set, the coupling to phonons is sufficiently strong to make the original LR scheme the optimal one over the whole range of coherence times.

\section{Conclusions}

We theoretically compared four schemes for the generation of linear cluster states with a charged semiconductor quantum dot: the original LR scheme, a modified version using time-bin encoding and two schemes using a cavity-induced cycling transition and optical spin control, either via a single ultrafast pulse or via Raman pulses. We find that the original LR scheme excels for long spin coherence times and benefits strongly from enhancement of the trion relaxation rate. Conversely, the scheme using a cavity-induced cycling transition and ultrafast control pulses becomes the optimal choice for lower coherence times and does not require strong enhancement of the trion relaxation for optimal fidelities, but it strongly benefits from suppression of the non-enhanced transitions. Interestingly, we find that the LR-scheme, using polarization encoding, is naturally robust against phonon-induced decoherence during the trion excitation step. For finite coherence times, an emitter with two orthogonally polarized cycling transitions, enabling polarization encoding, and an additional excited state coupling to both ground states, simultaneously allowing for optical spin control, might perform better than the present schemes, but we are unaware of such a level scheme existing in single quantum dots.

Our model includes a detailed description of the electronic QD states, the driving pulses, the cavity and the phonon bath. Combined with realistic parameters, this enables our investigations to accurately model and guide experiments. A logical next step would be to combine our model with a more detailed description of the decoherence affecting the hole spin. This could be used to include the magnetic field dependence of the coherence time and to investigate different schemes used to extend the coherence time. Alternatively, the model could be extended by incorporating more electronic states, for example light holes, which can mix with the heavy holes and affect the optical selection rules. This would likely degrade the fidelity of the present schemes, but could also be exploited to allow for optical spin control combined with robust cycling transitions \cite{koong2026coherent}.

Natural ways to expand upon this work would be the investigation of other quantum emitter systems like quantum dot molecules \cite{vezvaee2022deterministic, bopp2023coherent}, atomic emitters \cite{thomas2022efficient, yang2022sequential} or color centers \cite{tchebotareva2019entanglement} in a similar way. Coupling emitters capable of generating linear cluster states via controlled phase gates would allow for the generation of more complex graph states, like graph repeater states, ladder cluster stater states, or other larger and higher dimensional graph and cluster states \cite{buterakos2017deterministic, russo2018photonic, economou2010optically, gimeno2019deterministic}. A useful near-term objective would be identifying a suitable emitter system and generation scheme for the generation of small two-dimensional cluster states. Additionally, suitable systems with $d$ ground states would also allow for the generation of qudit graph states \cite{raissi2024deterministic}.

\section*{Acknowledgements}
The work was supported by the Deutsche Forschungsgemeinschaft (German Research Foundation) through the transregional collaborative research center TRR~142/3-2022 (231447078, project C09) and by the Paderborn Center for Parallel Computing, PC$^2$.

\appendix

\section{System Hamiltonian and Equations of Motion}\label{sec:hamiltonian}

In this section, we describe the Hamiltonian of our model system and the corresponding equations of motion. The total Hamiltonian $H$ is the sum of the following terms. As depicted in figure \ref{fig:temp1} (a) and (b), we consider two hole states and two trion states with a trion energy $E_\mathrm{trion}$
\begin{equation}
    H_\mathrm{dot} = E_\mathrm{trion} (\ket{T_\uparrow}\bra{T_\uparrow}+\ket{T_\downarrow}\bra{T_\downarrow}).
\end{equation}
The cavity is modeled as two orthogonally polarized, possibly degenerate photon modes
\begin{equation}
    H_\mathrm{cav} = \hbar\omega_H a_H^\dagger a_H+\hbar\omega_V a_V^\dagger a_V,
\end{equation}
where $\hbar\omega_H$ and $\hbar\omega_V$ are the energies of the modes and $a_H^{(\dagger)}$ and $a_V^{(\dagger)}$ are the corresponding annihilation (creation) operators. The dot-cavity interaction is included via a Jaynes-Cummings-like term
\begin{equation}
    H_\mathrm{dot-cavity} = g( \ket{T_\uparrow}\bra{\Uparrow} a_{\sigma_-} + \ket{T_\downarrow}\bra{\Downarrow} a_{\sigma_+} ) + \mathrm{h.c.},
\end{equation}
where $g$ denotes the coupling strength, h.c. denotes the Hermitian conjugate and $a_{\sigma_\pm}^{\dagger}=(a_H^{\dagger}\pm \mathrm{i} a_V^{\dagger})/\sqrt{2}$ relates the photonic creation operators for the different polarization bases. The in-plane magnetic field is described by
\begin{equation}
    H_B = \frac{\mu_B B_x}{2}(g_\mathrm{hole}\ket{\Uparrow}\bra{\Downarrow} + g_\mathrm{trion}\ket{T_\uparrow}\bra{T_\downarrow}) + \mathrm{h.c.},
\end{equation}
where $\mu_B$ denotes the Bohr magneton, $B_x$ is the field strength and $g_\mathrm{hole}$ and $g_\mathrm{trion}$ are the Landé-factors of the hole and the trion. The driving laser pulses are described in rotating wave and dipole approximation by
\begin{equation}
    \begin{split}
        H_\mathrm{dot-laser} = -\frac{\hbar}{2}\sum_k \Omega_k(t) ( &\ket{T_\uparrow}\bra{\Uparrow} \vec{e}_k \cdot \vec{e}_{\sigma_-} \\
        + &\ket{T_\downarrow}\bra{\Downarrow} \vec{e}_k \cdot \vec{e}_{\sigma_+} ) + \mathrm{h.c.},
    \end{split}
\end{equation}
where $\vec{e}_{\sigma_+}=(1,0)^\mathrm{T}$ and $\vec{e}_{\sigma_-}=(0,1)^\mathrm{T}$ are the circular polarization basis vectors and $\vec{e}_k$, with $|\vec{e}_k|=1$, is the polarization of the $k^\mathrm{th}$ pulse in the circularly polarized basis. Each pulse is given by
\begin{equation}
    \Omega_k(t) = \frac{\alpha_k}{\pi}\sigma_k \mathrm{sech}\left(\frac{t-t_{0,k}}{\sigma_k}\right) e^{-i \omega_k t + i \phi_k}
\end{equation}
with the pulse area $\alpha_k$, the temporal width $\sigma_k$, the frequency $\omega_k$ and the phase $\phi_k$ and is centered around $t_{0,k}$.

For the calculations without phonons the system dynamics are obtained by solving the Lindblad equation
\begin{equation}
    \frac{d}{dt}\rho = -\frac{i}{\hbar} [H, \rho] + \mathcal{L} \rho
\end{equation}
for an initial condition $\rho(t_0)=\rho_0$ where
\begin{equation}
    \mathcal{L}\rho = \sum_i A_i \rho A_i^\dagger - \frac{1}{2}(A_i^\dagger A_i \rho + \rho A_i^\dagger A_i)
\end{equation}
and each operator $A_i$ describes a decay or dephasing channel. We include the cavity emission with $A_{\kappa,\sigma_+}=\sqrt{\kappa} a_{\sigma_+}$ and $A_{\kappa,\sigma_-}=\sqrt{\kappa} a_{\sigma_-}$, the non-cavity/radiative decay with $A_{\mathrm{rad},\uparrow}=\sqrt{\gamma_\mathrm{rad}}\ket{\Uparrow}\bra{T_\uparrow}$ and $A_{\mathrm{rad},\downarrow}=\sqrt{\gamma_\mathrm{rad}}\ket{\Downarrow}\bra{T_\downarrow}$ and the dephasing of the ground state qubit with $A_{\mathrm{deph}, \Uparrow}=\sqrt{\gamma_\mathrm{deph}}\ket{\Uparrow}\bra{\Uparrow}$ and $A_{\mathrm{deph}, \Downarrow}=\sqrt{\gamma_\mathrm{deph}}\ket{\Downarrow}\bra{\Downarrow}$. We numerically solve the Lindblad equation in matrix representation in the interaction picture with a fourth-fifth-order adaptive-stepsize Runge-Kutta scheme. The multi-time expectation values required for the evaluation of the photonic correlation functions are computed with the quantum regression theorem \cite{breuer2002theory}.

\section{Phonon Model}\label{sec:hamiltonian_phonons}

In section \ref{sec:phononInfluence} we investigate the influence of phonons on the state preparations schemes. We include longitudinal acoustic (LA) phonons and deformation potential coupling with
\begin{equation}
    \begin{split}
    H_\mathrm{phonon} = &\sum_\mathbf{q} \hbar\omega_\mathbf{q} b^\dagger_\mathbf{q} b_\mathbf{q} \\ 
    + &\sum_\mathbf{q} \hbar \lambda_\mathbf{q}(\ket{T_\uparrow}\bra{T_\uparrow}+\ket{T_\downarrow}\bra{T_\downarrow})(b_\mathbf{q}+b^\dagger_\mathbf{q}),
    \end{split}
\end{equation}
where $b_\mathbf{q}^{(\dagger)}$ annihilates (creates) a phonon with wave vector $\mathbf{q}$ and $\omega_\mathbf{q}$ and $\lambda_\mathbf{q}$ describe the frequency of the phonon modes and their coupling to the electronic states. The phonons are characterized by their spectral density
\begin{equation}
    J(\omega) = \alpha_p \omega^3 \exp(-\frac{\omega^2}{2 \omega_b^2}),
\end{equation}
where $\alpha_p$ is the phonon coupling strength and ${\hbar\omega_b=\SI{1}{\meV}}$ is the phonon cutoff frequency. To compute the system dynamics for the case with phonons, we use a process tensor matrix product operator method, which allows for a numerically exact treatment of the phonon environment \cite{cygorek2022simulation}. For our computations we use the solver provided in reference \cite{githubACE} with the periodic PT algorithm described in reference \cite{cygorek2024sublinear}. This approach allows for the evaluation of multi-time expectation values in the presence of phonons without relying on the quantum regression theorem.

\section{Photonic Correlation Functions}\label{sec:correlationFunctionsExamples}

In this section we show explicit examples of the correlation functions from section \ref{sec:correlationFunctions} for the different generation schemes. As an example we use $\bar{g}^{(3)}_{011100}$, required for the calculation of $\langle Z^{(1)} X^{(2)} Z^{(3)} \rangle$. The original LR scheme uses polarization encoding and the time-bins, each encompassing a full photonic qubit, have a duration of a quarter spin precession time $T_B/4$. With the encoding from section \ref{sec:correlationFunctions} the corresponding correlation function reads
\begin{equation}
    \begin{split}
        &\bar{g}^{(3)}_{011100} = \bar{g}^{(3)}_{\sigma_+,\sigma_-,\sigma_-,\sigma_-,\sigma_+,\sigma_+} \\
        =& \int_{\frac{2}{4} T_B/2}^{\frac{3}{4} T_B} \int_{\frac{1}{4} T_B/2}^{\frac{2}{4} T_B} \int_{0}^{\frac{1}{4} T_B} dt_1 dt_2 dt_3 \\
        &\times \langle a^\dagger_{\sigma_+}(t_1) a^\dagger_{\sigma_-}(t_2) a^\dagger_{\sigma_-}(t_3) a_{\sigma_-}(t_3) a_{\sigma_+}(t_2) a_{\sigma_+}(t_1) \rangle.
    \end{split}
\end{equation}
For the optically controlled schemes using time-bin encoding, each photonic qubit encompasses an early and a late time-bin of length $T$, resulting in
\begin{equation}
    \begin{split}
        &\bar{g}^{(3)}_{011100} = \bar{g}^{(3)}_{ELLLEE} \\
        =& \int_{4T}^{5T} \int_{2T}^{3T} \int_0^T dt_1 dt_2 dt_3 \\
        &\times \langle a^\dagger_E(t_1) a^\dagger_L(t_2) a^\dagger_L(t_3) a_L(t_3) a_E(t_2) a_E(t_1) \rangle \\
        =& \int_{4T}^{5T} \int_{2T}^{3T} \int_0^T dt_1 dt_2 dt_3 \\
        &\times \langle a^\dagger(t_1) a^\dagger(t_2+T) a^\dagger(t_3+T) a(t_3+T) a(t_2) a(t_1) \rangle.
    \end{split}
\end{equation}
Here, all creation and annihilation operators refer to the horizontally polarized cavity mode ($a^{(\dagger)}_H$), but for clarity we omit the index $H$. For the time-bin LR scheme this correlation function needs to be modified to account for the unequal delays between the excitation pulses. The scheme requires alternating between $R_x(\pi)$ and $R_x(\pi/2)$ gates, corresponding to delays of $T_B/2$ and $T_B/4$. Therefor the excitation pulse for the $n^\mathrm{th}$ early bin is applied shortly after the beginning of the bin ${t=(n-1)\frac{3T_B}{4}}$ and the excitation pulse for the corresponding late bin is delayed by $T_B/2$. The resulting correlation function reads
\begin{equation}
    \begin{split}
        &\bar{g}^{(3)}_{011100} = \bar{g}^{(3)}_{ELLLEE} \\
        =& \int_{\frac{6}{4}T_B}^{\frac{7}{4}T_B} \int_{\frac{3}{4}T_B}^{\frac{4}{4}T_B} \int_{0}^{\frac{1}{4}T_B} dt_1 dt_2 dt_3 \\
        &\times \langle a^\dagger(t_1) a^\dagger(t_2+{\textstyle\frac{T_B}{2}}) a^\dagger(t_3+{\textstyle\frac{T_B}{2}}) a(t_3+{\textstyle\frac{T_B}{2}}) a(t_2) a(t_1) \rangle.
    \end{split}
\end{equation}
Here all creation and annihilation operators refer to the $\sigma_-$-polarized mode ($a^{(\dagger)}_{\sigma_-}$) and we again omit the indices for clarity.

While we have only shown an example of a single third order correlation function for each scheme, the $N$-photon correlation functions for other combinations of indices and for different numbers of photons behave similarly.

\section{Lower Bound on Entanglement Length}\label{sec:lowerBound}

Here we derive the lower bound given in equation (\ref{eq:entanglementLength}). We start with a witness detecting genuine $N$-partite entanglement in $N$-qubit graph states \cite{toth2005entanglement}
\begin{equation}
    \mathcal{W}^{(N)} = (N-1)\mathds{1} - \sum_{i=1}^N S_i,
\end{equation}
where $S_i$ are the stabilizer generators of the graph state and $\mathds{1}$ is the $N$-qubit identity operator. For a linear cluster state, assuming all stabilizer generators except for $S_1$ have the same expectation value, the expectation value of the witness reduces to
\begin{equation}
    \langle \mathcal{W}^{(N)} \rangle = (N-1) (1-\langle Z^{(1)} X^{(2)} Z^{(3)} \rangle) - \langle X^{(1)} Z^{(2)} \rangle. 
\end{equation}
This is justified, even for the last stabilizer generator $S_N$, as measuring $S_N=Z^{(N-1)}Z^{(N)}$ and decoupling the emitter via a measurement of $Z^{(N+1)}$ is equivalent to measuring $Z^{(N-1)} X^{(N)} Z^{(N+1)}$. Demanding $\langle \mathcal{W}^{(N)} \rangle < 0$ results in a condition for $N$, for which genuine multipartite entanglement can be detected
\begin{equation}
    N < \frac{\langle X^{(1)} Z^{(2)} \rangle}{1- \langle Z^{(1)} X^{(2)} Z^{(3)}\rangle} + 1
\end{equation}
and thus in the lower bound described in the main text. There exist witnesses, that can detect cluster state entanglement in a larger class of states, but they require the evaluation of products of stabilizer generators and thus higher order correlation functions, which are significantly more computationally expensive \cite{toth2005entanglement}.

\section{Phonon-Induced Decoherence with Polarization Encoding}\label{sec:phononProtectionLR}

In this section, we give a qualitative explanation as to why the fidelity of cluster states generated with the original Lindner-Rudolph scheme is nearly unaffected by coupling to phonons. We assume that the state of the full system can be written as 
\begin{equation}
    \ket{\Psi_\mathrm{dot}}\ket{\Phi_\mathrm{photons}}\ket{\chi_\mathrm{phonons}}
\end{equation}
or as a linear combination of such product states, where $\ket{\Psi_\mathrm{dot}}$ describes the state of the QD, $\ket{\Phi_\mathrm{photons}}$ describes the state of the emitted photons and $\ket{\chi_\mathrm{phonons}}$ describes the state of the phononic environment. We note that explicitly including a cavity does not affect our argument. Here, we explicitly use the states involved in the spin-precession based scheme (z-basis), but the argument does not depend on the exact choice of states. We initialize the system in the state
\begin{equation}
    \frac{1}{\sqrt{2}}\left(\ket{\Uparrow}\ket{\emptyset}\ket{\chi_0}+\ket{\Downarrow}\ket{\emptyset}\ket{\chi_0}\right),
\end{equation}
where $\ket{\emptyset}$ denotes the absence of emitted photons and $\ket{\chi_0}$ denotes the initial state of the phonons. Exciting the ${\ket{\Uparrow} \leftrightarrow \ket{T_\uparrow}}$ transition leads to a disturbance of the phononic state and the emission of a $\sigma_-$-photon once the trion has fully decayed, both conditioned on the state of the hole qubit
\begin{equation}
    \frac{1}{\sqrt{2}}\left(\ket{\Uparrow}\ket{1_{\sigma_-}}\ket{\chi'}+\ket{\Downarrow}\ket{\emptyset}\ket{\chi_0}\right).
\end{equation}
As $\ket{\chi'}\neq\ket{\chi_0}$, this state shows entanglement between the dot-photon subsystem and the phonon subsystem. Tracing out the phonons results in a mixed state, where the degree of remaining coherence is related to the magnitude of the overlap $\braket{\chi'}{\chi_0}$. Exciting both trions simultaneously instead results in
\begin{equation}
    \frac{1}{\sqrt{2}}\left(\ket{\Uparrow}\ket{1_{\sigma_-}}\ket{\chi'}+\ket{\Downarrow}\ket{1_{\sigma_+}}\ket{\chi''}\right).
\end{equation}
If the coupling of both trions to the phonons is equally strong, which is given for our system, we have $\ket{\chi'}=\ket{\chi''}$ and thus no entanglement with the phononic environment and no phonon-induced decoherence acting on the spin-photon cluster state. A similar effect can be observed for the generation of entangled photon pairs via the biexciton cascade, when both bright exciton states are degenerate \cite{carmele2011analytical, seidelmann2019strong}.

\bibliography{bibo}

\end{document}